# A high entropy alloy as very low melting point solder for advanced electronic packaging

Yingxia Liu[a*], Li Pu[a], Yong Yang[b,c*] Quanfeng He[b], Ziqing Zhou[b], Chengwen Tan[a], Xiuchen Zhao[a], Qingshan Zhang[a] and K. N. Tu[d]

a: Dept. of Materials Science and Engineering, Beijing Institute of Technology, Beijing, China

b: Dept. of Mechanical Engineering, City University of Hong Kong, Hong Kong, China

c: Dept. of Materials Science and Engineering, City University of Hong Kong, Hong Kong, China

d: Dept. of Materials Science and Engineering, University of California, Los Angeles, USA

*: yingxia.liu@bit.edu.cn

*: yonyang@cityu.edu.hk

**Abstract:** SnBiInZn based high entropy alloy (HEA) was studied as a low reflow temperature solder with melting point around 80 °C. The wetting angle is about 52° after reflow at 100 °C for 10 min. The interfacial intermetallic compound (IMC) growth kinetics was measured to be ripening-control with a low activation energy about 18.0 kJ/mol, however, the interfacial reaction rate is very slow, leading to the formation of a very thin IMC layer. The low melting point HEA solder has potential applications in advanced electronic packaging technology, especially for bio-medical devices.

1. **Introduction**

While Moore's law in Si chip technology is near ending, electronic packaging technology is becoming critically important in order to sustain the future computational growth in microelectronics industry. The trend in miniaturization of very-large-scale-integration (VLSI) is moving from 2D IC to 3D IC [1-3]. The latter has various chips stacking vertically, which requires the development of new technologies such as TSV (through-Si-Via) and micro-bumps. More importantly, the 3D IC packaging technology will need to use a hierarchy of solder joints. In other words, low (around 100 ºC), middle (200 ºC), and high (300 ºC) wetting temperature solders will work together, so that different components can be stacked and integrated. At the moment, we have the high-Pb $Pb_{95}Sn_5$ solder for the high melting point and the eutectic SnAg solder for the middle melting point [4-5]. But for the low melting point, we only have eutectic SnBi, which has a melting point of 138ºC with a soldering temperature about 150 ºC [6-7]. It would be better if we could lower the soldering temperature furthermore to 100 ºC. What's more, for the future bio-medical applications, it's important to have low reflow temperature solder applied to related packaging technologies, since the working temperature of bio-medical devices is body temperature. In this paper, we report an HEA solder with a melting point about 80 °C and a solder wetting temperature about 100 °C.

## 2. Experimental

SnBiInZn HEA solder were prepared using high purity (>99.9%) Sn, Bi, In and Zn as atomic ratio of Sn:Bi:In:Zn = 1:1:1:1 in a vacuum induction furnace. Pieces about 5-10 mg were cut from the bulk solder alloy. These pieces were analyzed by differential thermal analysis (DTA) to measure the melting point of SnBiInZn solder.

To study the wetting behavior, we polished copper foils with 2.5 μm diamond abrasion paste and cleaned with deionized water. Then a piece of 0.5 mg solder was placed on the copper foil merging in flux and reflowed on a hot-plate at 120 °C, 140 °C and 160 °C for 5 min, 10 min, and 20 min, respectively. We also succeeded to wet the HEA solder on a Cu plate at 100 °C for 10 min and 20 min. After reflow, we cooled the samples in air to room temperature and cleaned the samples with pure alcohol.

The wetting samples were mounted in epoxy resin. The cross-sections were polished with SiC papers successively and then with 0.04 μm $SiO_2$ powder suspension. We observed the cross-sections of the polished samples by scanning electron microscope (SEM). The elemental composition of IMC was analyzed by energy dispersive X-ray spectroscope (EDX). We measured the area and length of IMC by Image J. The thickness of IMC was obtained from area divided by length. Transmission electron microscope (TEM) images were acquired by FEI Themis Z FEG/TEM operated at 200 kV in Bright-Field (BF) Scanning Tunneling Electron Microscopy (STEM) mode and by high-angle annular dark-field (HAADF) STEM mode for more detailed information. High Resolution TEM (HRTEM) images are also acquired.

For shear tests, we reflowed 5 ± 0.5 mg of diced solder pieces on 1mm diameter circular Cu substrate at 100 °C to 160 °C for 1 min and 5 min. The shear tests were performed using PTR-1100 shear test machine at room temperature with a shear strain rate of 0.5 mm/s.

## 3. Results and Discussions

### 3.1 Interfacial microstructure and HEA solder matrix

SEM cross-sectional image of original HEA solder is shown in Fig. 1(a). There are three phases observed in Fig. 1(a): the Sn-rich phase, InBi phase and Bi phase. Fig. 1(b) and 1(c) are the XRD results of the original HEA solder and pure Sn, respectively. By comparison, we can see the Sn-rich phase in the alloy has broader peaks than pure Sn, also there is a tiny shift in the peak position, implying the Sn-rich phase is a solid solution. It needs to be noted that, we do have InBi and Bi phase in the alloy and the entropy of this alloy may not be as high as other HEAs. But the Sn-rich phase in the alloy does have a solid solution structure and for that phase, it has a high entropy. The HEA solder structure seems to be very stable and there is no big difference before and after reflow. More characterization results about the Sn-rich solid solution phase after reflow reaction will be presented.

DTA was performed to find the solder melting point around 80 °C, as shown in Fig

2(a). After being reflowed at 160 °C for 5 min and 20 min, another small peak around 56 °C occurred in the DTA curve, shown in Fig. 2(b) and (c), indicating that some phase transformation had taken place inside the HEA material during the reflow. Fig. 3(a) to (d) show the wetting angle after soldering for 10 min at 100 °C to 160 °C. The wetting angle is about 35 to 40 degrees after soldering at 120 °C to 160 °C for different length of time, but more than 50 degrees at 100 °C. When we further reduced the reflow time to 1 min, we noticed that at the reflow temperature of 140 °C and 160 °C, we can have successful soldering. However, we were unable to achieve repetitively good solder joint at the reflow temperature of 100 °C and 120 °C with just 1 min reflow time. Often, we obtained a very thin layer of IMC (less than 200 nm) in the solder joints after reflow. We tend to believe that the flux is not quite efficient at such a low working temperature (100 °C and 120 °C). Our future work would include the finding of a flux with a low working temperature at 100 °C for 1 min reflow.

The microstructure of the HEA cap, as well as the interfacial structure between the HEA and Cu was observed by SEM cross-sectional images and FIB. Fig. 4(a) and (b) are respectively the lower (1000X) and the higher (6000X) magnification cross-sectional SEM images of the solder after reflowing at 100 °C for 20 min, and the uniform HEA solder microstructure can be observed. Fig. 4(c) shows the cross-sectional FIB ion beam image of the solder after reflowing for 5 min at 160 °C. Fig. 4(d) shows the SEM cross-sectional image for the solder after reflowing for 20 min at 160 °C. According to the EDX results, shown in Fig 4(e), the IMC formed during soldering reaction is believed to be $Cu_6Sn_5$ with a few percent of In substituting Sn atoms. The microstructure of the HEA cap is very complicated and has at least three detectable phases, including the phases of almost pure Bi phase, InBi phase, and Sn-rich phase, as marked in both Fig. 4(b) and Fig. 4(d). These three phases have also been confirmed by X-ray diffraction (XRD). In the ion beam image in Fig. 4(c), more information could be revealed, where three to four different phases with different extent of gray could be seen. Some twin structure is observed in one of those phases, which is marked by the white arrow. In the IMC part, there seems to be two layers of IMCs, marked by the two black arrows.

To have a better understanding of the HEA solder matrix and the formed IMC after reaction, TEM images in Fig. 5 were obtained after the solder being reflowed for 5 min at 160 °C. Fig. 5(a) is a bright-field (BF) STEM image, and we can observe solder layer, IMC layer and Cu layer in the image. Fig. 5(b) is a higher magnification BF TEM image, and two layers of IMC can be distinguished. Fig. 5(c) is a HAADF STEM image and we can observe some Kirkendall voids located between IMC and the Cu substrate. In the IMC layer, there are some tiny darker spots. From the EDX results, shown in Fig. 5(d), where we did EDX line scan along the arrow in Fig. 5(a), we tend to regard those two layers as $Cu_3Sn$ and $Cu_6Sn_5$ with a few percent of Zn and In. The results consist with the EDX results obtained from SEM. Fig. 5(e) to (h) are to show the element Cu, In, Sn, and Zn element distribution and the EDX mapping area is indicated in the white rectangular in Fig. 5(a). As shown in Fig. 5(h), Zn element appears to be particles in the EDX mapping image, thus, those tiny spots in IMC might be Zn particles.

Fig. 6(a) is the same as Fig. 5(a) used to mark the detected locations. The diffraction pattern of the solder layer with zone axis of [1 0 1] is shown in Fig. 6(b). The solder matrix has body-centered tetragonal (bct) structure with measured lattice parameters a=b=0.676 nm, c=0.339 nm. Compared with Sn bct crystal structure, a=b=0.583 nm, c=0.318 nm, the measured lattice constants are about 10% distorted from pure Sn. The reason should be explained by the 30 at. % In and a few percent of Zn atoms in the lattice. The atomic radius of Sn, In and Zn is 145 pm, 155 pm and 135 pm, respectively [8]. We also obtained HRTEM images to show the lattice sites, as shown in Fig. 6(c), (d) and (e). Those images are acquired in locations marked in the red rectangular shown in Fig. 6(a). The three locations are around a hole, induced during the thinning when we make the FIB-TEM sample. The three locations all have the same crystal structure as pure Sn. According to Fig. 6, the solder matrix seems to have relatively large grains and each grain is a single crystal. We note that the solder composes of around 40 at. % In and Zn atoms, however, it still has perfect bct crystal structure. That is why we believe our solder is in high entropy state. Though the solder may not be defined exactly as HEA, since it has less than five main elements in it, it should be appropriate to define the solder as medium entropy alloy. The application of HEA as solder in this work is a novel try and should have plenty of following work in the future.

### *3.2 Growth kinetics of IMCs*

To study IMC growth kinetics, the thickness of the IMC formed at 120 ºC, 140 ºC and 160 ºC after different reflow time was measured. The measured thicknesses are plotted in Fig. 7(a). The mean thicknesses, L, can be described very well by Eq. (1), and the growth rate $D$ can be calculated using the following equation [4].

$$L = Dt^n \tag{1}$$

The Arrhenius relationship can be applied to obtain the activation energy, in the form below [4],

$$D = A\exp\left(-\frac{E_a}{RT}\right) \tag{2}$$

where $n$ is a reaction constant and $A$ is a pre-factor, $E_a$ is the activation energy, $R$ is the ideal gas constant, and $T$ is the absolute temperature. Fitting the measured thickness into Eq. (1) by taking the logarithm on both sides, we obtain the calculated $n$ to be 0.30, 0.32, and 0.36 respectively for 120, 140, and 160 ºC. The data fits well with the published $Cu_6Sn_5$ ripening growth kinetics data with $n = 1/3$ [9-10]. The activation energy is then calculated from the slopes of the fitted lines to be 18.0 kJ/mol, as shown in Fig. 7(b). Some published data on the activation energy for solid-liquid interfacial reactions of Cu-Sn has been reported to be 19.72 kJ/mol [11] and 29 kJ/mol [12]. The activation energy for Sn58Bi solder reaction with Cu is reported to be 17.6 kJ/mol [4], and eutectic SnPb solder reaction with Cu is 18.3 to 27.9 kJ/mol [9]. By comparison, we can figure that the activation energy we measured in HEA solder/Cu reaction is in the same range.

Significantly, while the activation energy of solid-liquid interfacial reaction of the HEA solder on Cu is in the same range as other conventional solders, the rate of IMC formation is much slower. It's worth noting that after HEA solder reflowing for 10 and 20 min at 100 ºC, the average IMC thickness is measured to be 1.32 and 1.49 μm, respectively. This is surprising and it could be a unique nature of the HEA alloy because the entropy factor is in the pre-factor "A" as shown in Eq. (2). In the temperature range of 100 to 160 °C, the solid-solid interfacial reaction occurs in the conventional solders. On the other hand, if we conduct solid-liquid interfacial reaction for 10 to 20 min, say in eutectic SnAg solder on Cu, the IMC will be over 10 μm [13-15], which is much thicker than that observed here. In theory, the pre-factor $A \propto \exp\left(\frac{\Delta S}{R}\right)$ and $S$ represents the activation entropy. Neglecting the change in vibrational entropy during the IMC formation, we may approximately take $\Delta S \approx S_{\text{tran}} - S_{\text{HEA}}$, where $S_{\text{tran}}$ and $S_{\text{HEA}}$ stand for the configurational entropy of the transition state and the original HEA respectively during IMC formation. Since $S_{\text{HEA}}$ of HEAs is very high at a high homologous temperature, which could reach the prediction of the ideal mixing rule [13-14], we herein propose that the low pre-factor $A$ may be attributed to the entropy reduction during the IMC formation in the HEA. In case that $S_{\text{tran}} \ll S_{\text{HEA}}$, $A \propto \exp\left(-\frac{S_{\text{HEA}}}{R}\right)$ and hence, high configurational entropy could lead to rather low reaction kinetics, as seen in our experiments.

It should be noted that, even after 20 min reflow at 100 °C, the IMC thickness is still very thin with an average thickness of 1.49 um. If we could find a high efficiency flux and obtain a successful wetting solder joint at 100 °C, we expect a very thin layer of IMC in the joint after reflow for 1 min. The thin thickness of IMC layer could relieve many yield and reliability issues in the future small size solder joints. For example, we might not need the Ni layer as the diffusion barrier on the Cu under bump metallization (UBM) to prevent the high Cu consumption rate as well as Kirkendall void formation.

The latter has been associated to the growth of a thick layer of $Cu_3Sn$ between $Cu_6Sn_5$ and Cu [15-17]. This is because the void nucleation requires the super-saturation of vacancies in $Cu_3Sn$ and in its interface with Cu. As shown in Fig. 5, there are only some tiny Kirkendall voids in our sample even after reflow for 5 min at 160 °C. Kirkendall voids related reliability issues would be mitigated by our solder.

*3.3 Shear test results*

The mechanical properties of this low melting point solder joint were investigated by shear test and the test results are listed in Table 1. The average shear strength is measured to be about 19 MPa to 28 MPa. According to the published data, solders with different composition, including Sn-0.4Cu, Sn-3Ag-0.4Cu, Sn-58Bi, and SnZnBi, have the shear strength of 19.5 MPa, 32.5 MPa, 64 MPa, and 18.5-28.0 MPa, respectively [18-20]. Comparing with the published data, we conclude that the HEA solder joint has a relatively good mechanical strength. Interestingly, even though at 100 °C and 120 °C, we can only have a very thin layer of IMC, the solder joint strength is still good.

|  | Temp (°C) | Shear stress first test (MPa) | Shear stress second test (MPa) | Average (MPa) |
|---|---|---|---|---|
| Reflow for 5min | 120 | 26.4 | 18.9 | 22.7 |
|  | 140 | 21.9 | 19.1 | 20.5 |
|  | 160 | 27.9 | 27.9 | 27.9 |
| Reflow for 1 | 100 | 27.6 | 24.6 | 26.1 |
|  | 120 | 25.7 | 24.7 | 25.2 |
| min | 140 | 18.3 | 19.9 | 19.1 |
| | 160 | 21.4 | 23.8 | 22.6 |

## 4. Conclusion

In summary, the HEA of SnBiInZn has been studied as a low melting point solder. It has good wetting properties and good shear strength at the reflow temperature of 100 °C. Moreover, the IMC growth kinetics study indicates that it has a very slow solid-liquid interfacial reaction rate during reflow, forming a very thin layer of $Cu_6Sn_5$ IMC. The reason was explained by the unique nature of the HEA alloy, because the pre-factor "*A*" in the Arrhenius relationship has the entropy factor. The application of HEA as low melting point solder in this work is a novel try and it has potential for applications in advanced electronic packaging technology in the future.

**Acknowledgements**

The authors at Beijing Institute of Technology would like to acknowledge the financial support from Youth Program of National Natural Science Foundation of China with the project number 51901022. The research of YY is supported by City University of Hong Kong with the project number 9610391.

Reference

[1] Y. Liu, M. Li, D.W. Kim, S. Gu, K. Tu, Synergistic effect of electromigration and

Joule heating on system level weak-link failure in 2.5 D integrated circuits, J. Appl. Phys. 118 (2015) 135304.

[2] K. Tu, Y. Liu, M. Li, Effect of Joule heating and current crowding on electromigration in mobile technology, Appl. Phys. Rev. 4 (2017) 011101.

[3] Y. Liu, Y.C. Chu, K. Tu, Scaling effect of interfacial reaction on intermetallic compound formation in Sn/Cu pillar down to 1 μm diameter, Acta Mater. 117 (2016) 146-152.

[4] A. Kroupa, D. Andersson, N. Hoo, J. Pearce, A. Watson, A. Dinsdale, S. Mucklejohn, Current problems and possible solutions in high-temperature lead-free soldering, J. Mater. Eng. Perform, 21 (2012) 629-637.

[5] R. Khazaka, L. Mendizabal, D. Henry, R. Hanna, Survey of high-temperature reliability of power electronics packaging components, IEEE T. power Electr. 30 (2014) 2456-2464.

[6] J. Li, S. Mannan, M. Clode, D. Whalley, D. Hutt, Interfacial reactions between molten Sn–Bi–X solders and Cu substrates for liquid solder interconnects, Acta Mater. 54 (2006) 2907-2922.

[7] J. Li, S. Mannan, M. Clode, K. Chen, D. Whalley, C. Liu, D. Hutt, Comparison of interfacial reactions of Ni and Ni–P in extended contact with liquid Sn–Bi-based solders, Acta Mater. 55 (2007) 737-752.

[8] J.C. Slater, Atomic radii in crystals, J. Chem. Phys. 41 (1964) 3199-3204.

[9] H. Kim, K. Tu, Kinetic analysis of the soldering reaction between eutectic SnPb alloy and Cu accompanied by ripening, Phys. Rev. B: Condens. Matter. 53 (1996)

16027-16034.

[10] M.M. Salleh, S. McDonald, H. Yasuda, A. Sugiyama, K. Nogita, Rapid Cu6Sn5 growth at liquid Sn/solid Cu interfaces, Scripta Mater. 100 (2015) 17-20.

[11] J. Li, P. Agyakwa, C. Johnson, Interfacial reaction in Cu/Sn/Cu system during the transient liquid phase soldering process, Acta Mater. 59 (2011) 1198-1211.

[12] C. Kao, Microstructures developed in solid-liquid reactions: using Cu-Sn reaction, Ni-Bi reaction, and Cu-In reaction as examples, Mater. Sci. Eng. A. 238 (1997) 196-201.

[13] Y. Ye, Q. Wang, J. Lu, C. Liu, Y. Yang, High-entropy alloy: challenges and prospects, Mater. Today, 19 (2016) 349-362.

[14] Q. He, Z. Ding, Y. Ye, Y. Yang, Design of high-entropy alloy: a perspective from nonideal mixing, JOM, 69 (2017) 2092-2098.

[15] P.T. Vianco, A Review of Interface Microstructures in Electronic Packaging Applications: Soldering Technology, JOM, 71 (2019) 158-177.

[16] K. Zeng, R. Stierman, T.C. Chiu, D. Edwards, K. Ano, K. Tu, Kirkendall void formation in eutectic SnPb solder joints on bare Cu and its effect on joint reliability, J. Appl. Phys. 97 (2005) 024508.

[17] H.Y. Hsiao, C.M. Liu, H.w. Lin, T.C. Liu, C.L. Lu, Y.S. Huang, C. Chen, K. Tu, Unidirectional growth of microbumps on (111)-oriented and nanotwinned copper, Sci. 336 (2012) 1007-1010.

[18] Keller, D. Baither, U. Wilke, G. Schmitz, Mechanical properties of Pb-free SnAg solder joints, Acta Mater. 59 (2011) 2731-2741.

[19] O. Mokhtari, H. Nishikawa, Correlation between microstructure and mechanical properties of Sn–Bi–X solders, Mater. Sci. Eng. A. 651 (2016) 831-839.

[20] J. Zhou, Y. Sun, F. Xue, Properties of low melting point Sn–Zn–Bi solders, J. Alloys Compd. 397 (2005) 260-264.

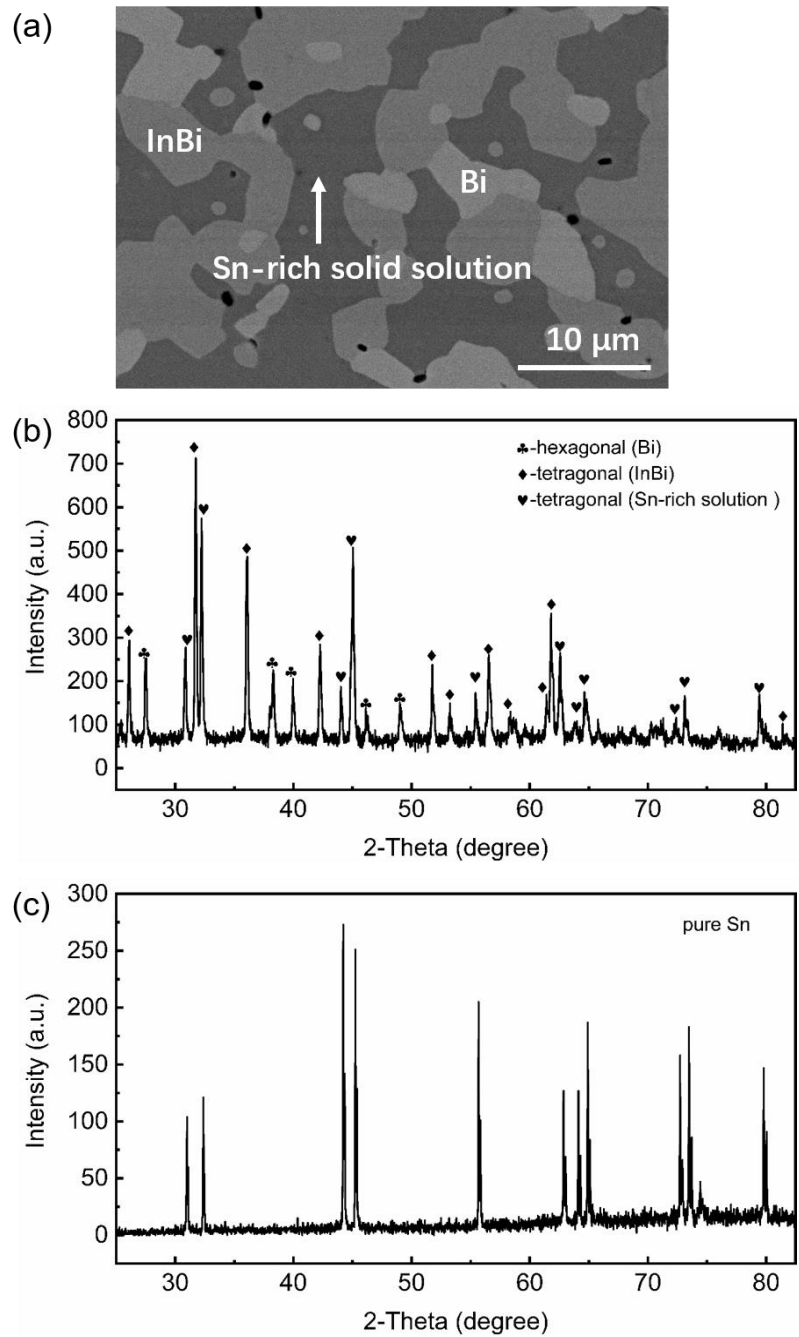

Fig. 1 (a) the SEM image of the original HEA solder. Fig. 1(b) and (c) XRD results of the HEA solder and pure Sn.

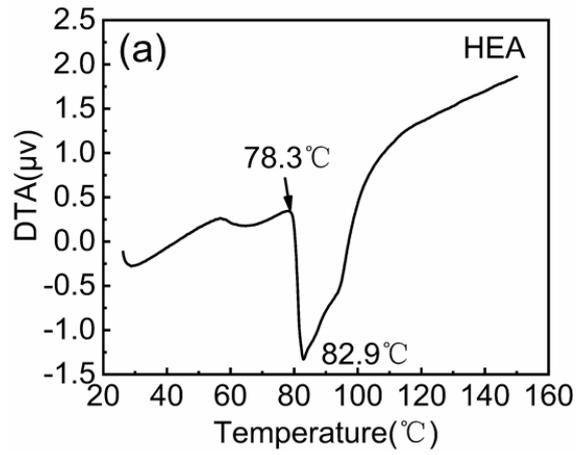

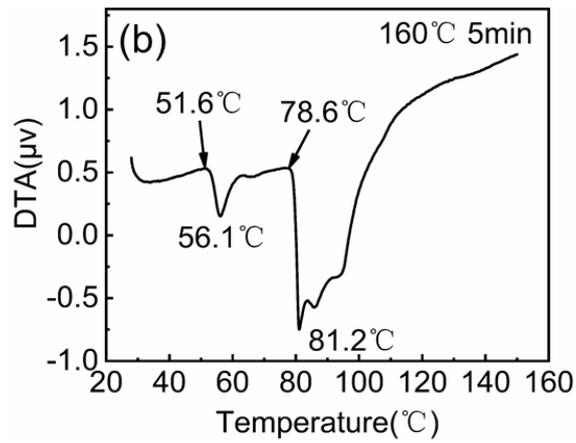

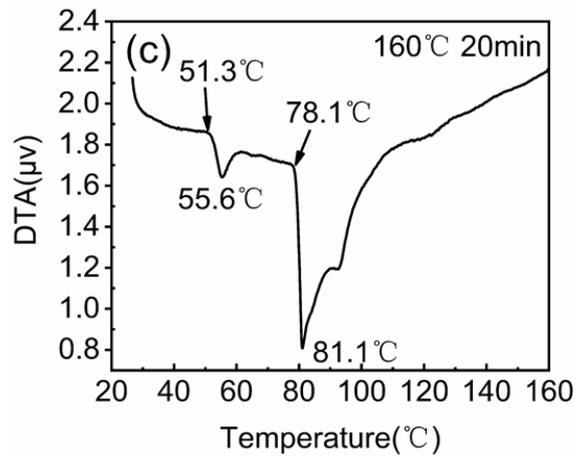

Fig. 2(a) DTA result for Original HEA material; Fig. 2(b) and (c) DTA result for HEA after being reflowed at 160 ºC for 5 min and 20 min.

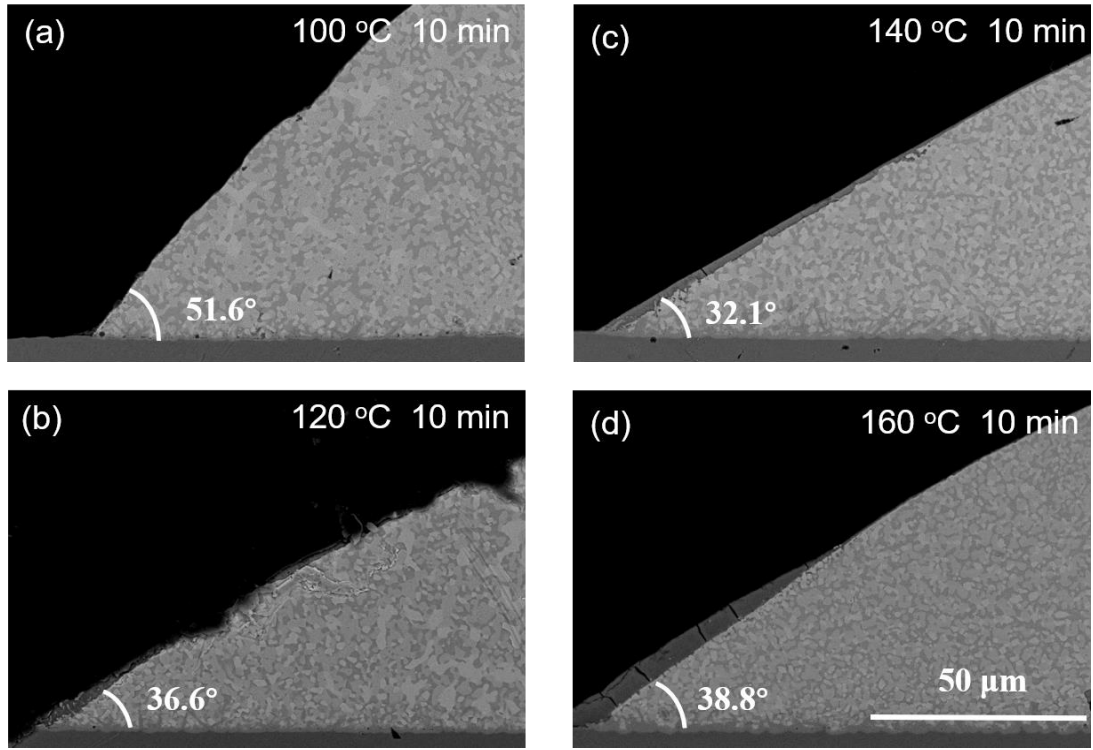

Fig. 3(a)-(d) The wetting angles after being reflowed for 10 min at different temperatures, (a) 100 °C; (b) 120 °C; (c) 140 °C; (d) 160 °C.

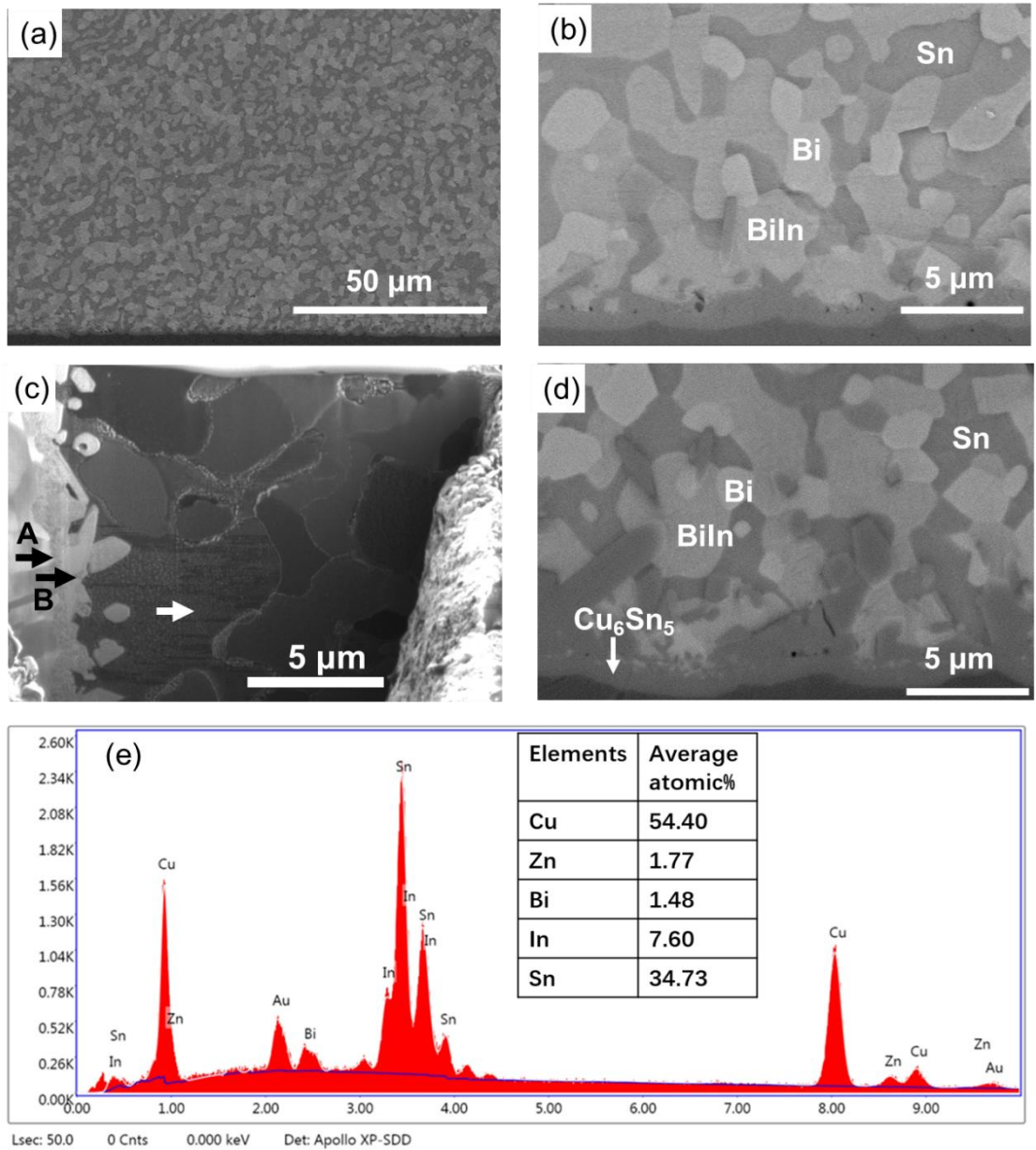

Fig. 4(a) and Fig. 4(b). The lower (1000X) and higher (6000X) magnification SEM images of the solder after reflowing at 100 °C for 20 min; Fig. 4(c). The FIB ion beam image of the solder after reflowing for 5 min at 160 °C; Fig. 4(d). The SEM image for the solder after reflowing for 20 min at 160 °C; Fig 4(e). the EDX result in the IMC area.

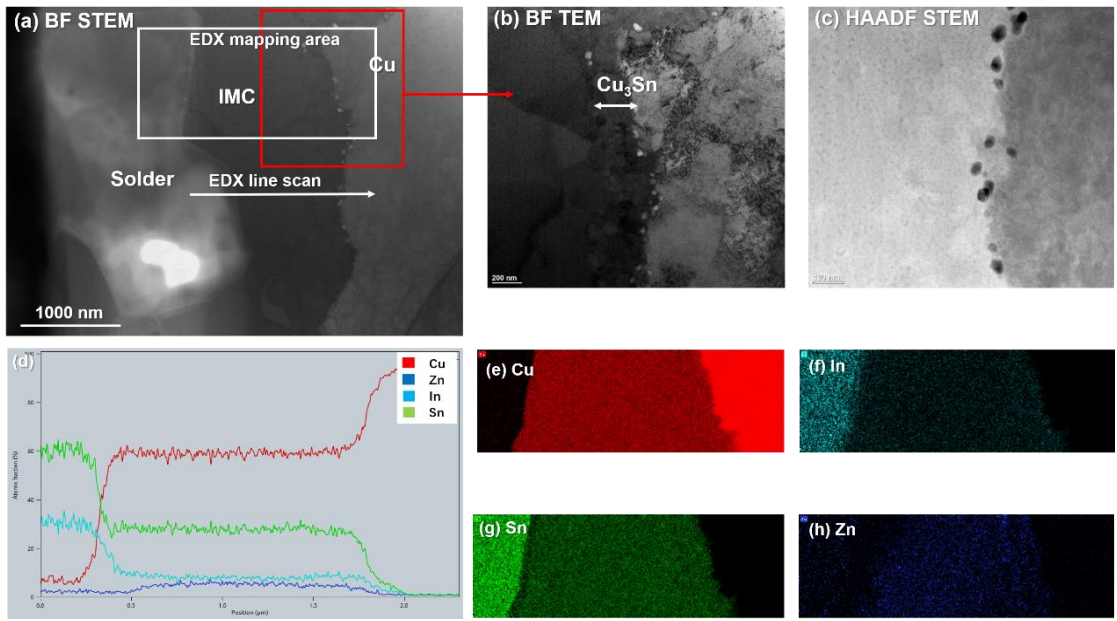

Fig. 5 TEM images after solder reflowing on Cu substrate at 160 ºC for 5 min; Fig. 5(a) BF STEM image for the interface; Fig. 5(b) Higher magnification BF TEM image; Fig. 5(c) HAADF STEM image; Fig. 5(d) EDX line scan results; Fig. 5(e), (f), (g) and (h) EDX mapping to show the distribution of element Cu, In, Sn, and Zn.

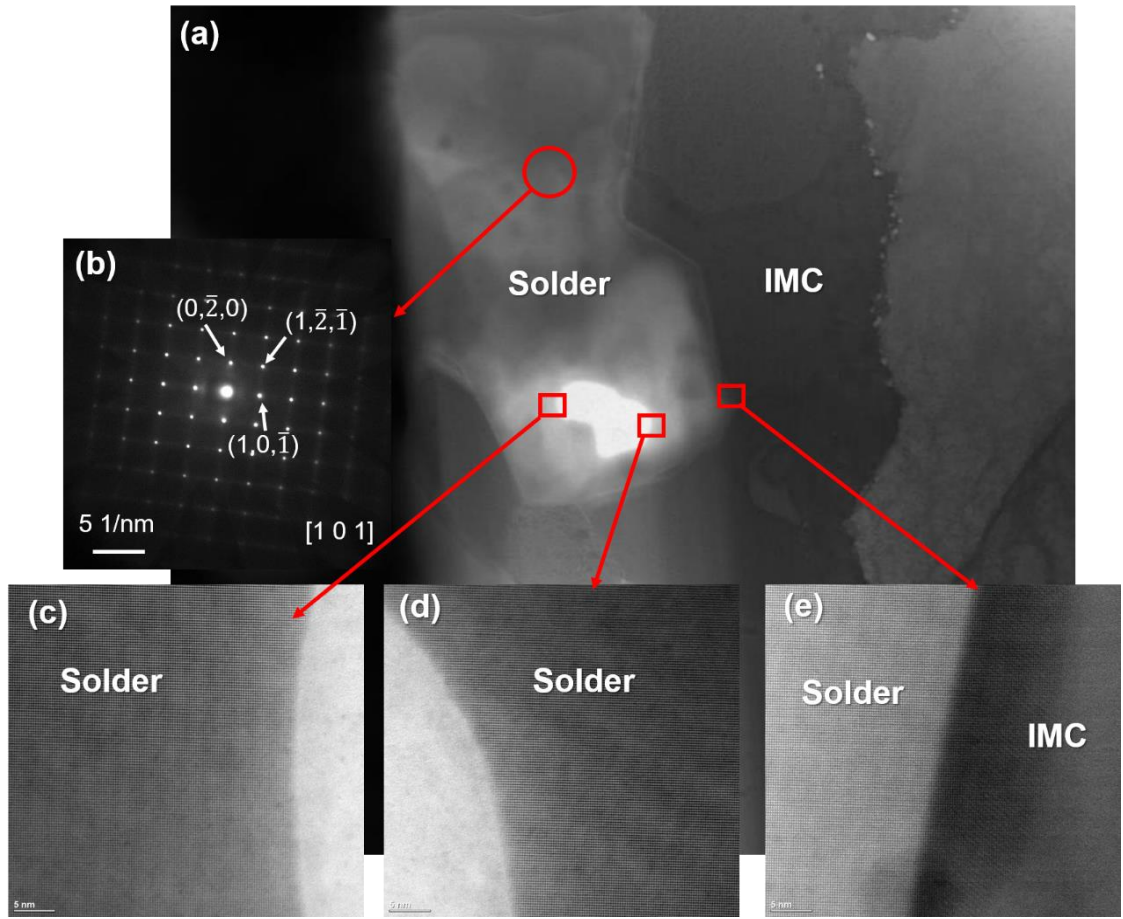

Fig. 6(a) BF STEM image for the interface; Fig. 6(b) The diffraction pattern for solder area; Fig. 6(c), (d) and (e) HRTEM images to show solder matrix lattice sites.

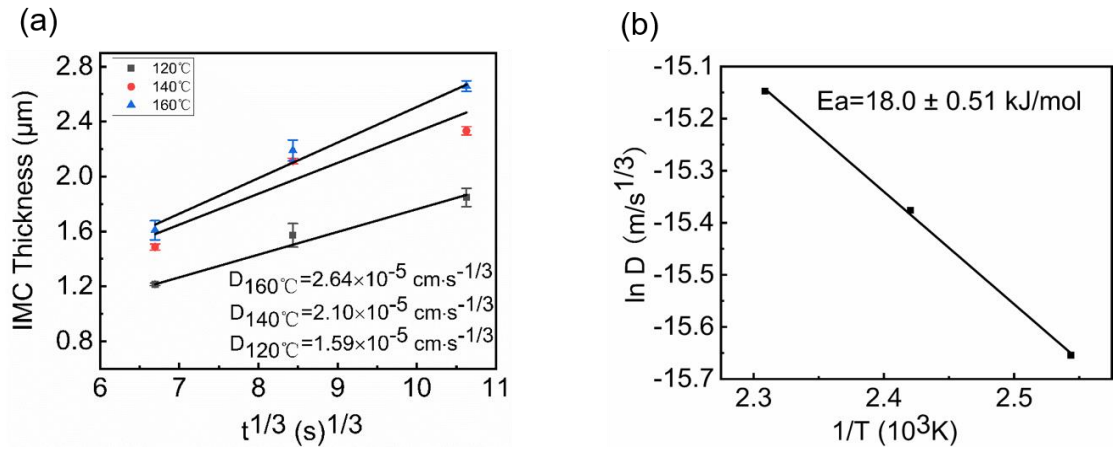

Fig. 7(a) Measured IMC thickness plotted with $t^{1/3}$; and Fig. 7(b) Arrhenius-type plot of the growth rate constant to 1/T.